\documentclass[nocopyrightspace]{sigplanconf}
\usepackage[nocompress]{cite} 
\usepackage{amsmath,amssymb, latexsym}

\usepackage{graphicx}
\usepackage[usenames,dvipsnames]{color}
\usepackage{enumerate}
\usepackage{hyperref}
\def\url{}
\usepackage{xspace}
\usepackage{epsfig}
\usepackage{mathpartir}
\usepackage{booktabs}
\usepackage{extarrows}
\usepackage{pgf,tikz}
\usetikzlibrary{arrows,automata}

\usepackage{ifthen}

\newcommand{\isTechReport}{true} 
\newcommand{\includeProof}[1]{
  \ifthenelse{\equal{\isTechReport}{true}}{
    #1
  }{
  }
}

\def\qed{\hfill$\Box$}

\def\@envspa{\hspace{0.3em}}
\def\@sa{\hspace{-0.2em}}
\def\@sb{\hspace{0.5em}}
\def\@sc{\hspace{-0.1em}}

{\itshape}{}
{\bfseries\upshape}{\itshape}
\newtheorem{@protheo}{Theorem}
\newenvironment{theorem}[1]{\begin{@protheo}{\rm \bf #1}\it}{\end{@protheo}}
{\bfseries\upshape}{\rm}
{\bfseries\upshape}{\itshape}
{\bfseries\upshape}{\itshape}
\newtheorem{lemma}{Lemma}{\bfseries\upshape}{\itshape}
{\bfseries\upshape}{\itshape}

\newcommand{\figbegin}{}
\newcommand{\figend}{}
\newcommand{\defeq}{\doteq\ }

\newcommand{\dbrkts}[1]{[\![#1]\!]}

\def\myexsh{\smallskip\noindent\textbf{\emph{Example.}\xspace}}

\def\mypara#1{\smallskip\noindent\textbf{#1}}

\def\set#1{{\{ #1\}}}

\def\tuple#1{{\langle #1 \rangle}}

\newcommand{\ie}{\textit{i.e.,}\xspace}
\newcommand{\eg}{\textit{e.g.,}\xspace}
\def\etc{{\it etc.}}

\newcommand{\translate}[1]{\dbrkts{#1}}
\newcommand{\Convert}{\alpha}
\newcommand{\Soln}{S}

\newcommand\Java{\textsc{Java}\xspace}
\newcommand\ML{\textsc{ML}\xspace}
\newcommand\HMC{\textsc{RTI}\xspace}

\newcommand\ocaml{\textsc{Ocaml}\xspace}

\newcommand\dsolve{\textsc{Dsolve}\xspace}

\newcommand\ARMC{\textsc{ARMC}\xspace}

\def\true{{\it true}}
\def\false{{\it false}}

\newcommand{\ttlength}{\mathtt{length}}
\newcommand{\ttf}{\mathtt{f}}
\newcommand{\ttg}{\mathtt{g}}

\newcommand{\ttxs}{\mathtt{xs}}
\newcommand{\ttxsp}{\mathtt{xs'}}

\newcommand{\ttitr}{\mathtt{itr}}
\newcommand{\ttvs}{\mathtt{vs}}
\newcommand{\ttgetValidSamples}{\mathtt{getValidSamples}}
\newcommand{\ttsumValidSamples}{\mathtt{sumValidSamples}}

\newcommand{\ttx}{\mathtt{x}}
\newcommand{\tty}{\mathtt{y}}

\newcommand{\ttb}{\mathtt{b}}
\newcommand{\tta}{\mathtt{a}}
\newcommand{\tti}{\mathtt{i}}
\newcommand{\ttj}{\mathtt{j}}

\newcommand{\ttList}{\mathtt{List}}
\newcommand{\ttget}{\mathtt{get}}
\newcommand{\ttset}{\mathtt{set}}

\newcommand{\ttassert}{\mathtt{assert}}

\newcommand{\ttjiter}{\mathtt{Iterator}}
\newcommand{\ttiteri}{\mathtt{iteri}}
\newcommand{\ttmask}{\mathtt{mask}}

\def\tte{\mathtt{e}}

\def\ttl{\mathtt{l}}

\def\ttxs{\mathtt{xs}}

\newcommand{\ttlenl}{{\ttlen}\xspace}
\newcommand{\ttlena}{{\ttlen}\xspace}

\def\zug#1{{\langle #1 \rangle}}

\makeatletter
\begingroup \catcode `|=0 \catcode `[= 1
\catcode`]=2 \catcode `\{=12 \catcode `\}=12
\catcode`\\=12 |gdef|@xcomment#1\end{comment}[|end[comment]]
|endgroup
\def\@comment{\let\do\@makeother \dospecials\catcode`\^^M=10\def\par{}}
\def\begincomment{\@comment\@xcomment}
\makeatother
\newenvironment{comment}{\begincomment}{}

\newcommand{\htab}{\hspace{0.4cm}}

\newcommand{\incode}[1]{\verb|#1|}  

\def\subt{<:}
\def\valu{\nu}

\newcommand{\ftyp}[2]{{{#1}\!:\!{#2}}}
\newcommand{\SUBST}[3]{{#1}[{#3}/{#2}]}

\newcommand{\EXT}[2]{{#1;\!#2}}

\newcommand{\UPD}[3]{{#1}[{#2} \mapsto {#3}]}

\newcommand{\ASSERT}[2]{\{{#1}\}\ {#2}}

\newcommand{\ttif}{\mathtt{if}}

\newcommand{\ttarray}{\xspace \mathtt{array}}
\newcommand{\ttfarray}{\xspace \mathtt{float[]}}
\newcommand{\ttlist}{\ \mathtt{list}}

\newcommand{\hastype}{::}

\newcommand{\mybar}[1]{\zug{#1}}

\newcommand{\ERROR}{\mathit{Error}}

\newcommand{\kvar}{\kappa}

\newcommand{\tenv}{\Gamma}
\newcommand{\renv}{G}
\newcommand{\univ}[1]{\mathcal{U}({#1})}

\def\ttunint{\mathtt{ui}}
\def\ttint{\mathtt{int}}
\def\ttbool{\mathtt{bool}}

\newcommand{\typ}{\tau}

\newcommand{\tliqs}{T}

\newcommand{\shape}[1]{\mathsf{Shape}({#1})}
\newcommand{\embed}[1]{\mathsf{Emb}({#1})}

\newcommand{\ttunit}{\mathtt{unit}}

\newcommand{\reftyp}[3]{\set{\ftyp{{#1}}{{#2}} \mid {#3}}}
\newcommand{\sreftyp}[1]{\set{{#1}}}
\newcommand{\deriv}{\vdash\ }

\newcommand{\rela}{\bowtie}

\newcommand{\ttlen}{\mathtt{len}}

\def\NONDET{\mathtt{nondet}}
\newcommand{\ilang}{\textsc{Imp}\xspace}
\newcommand{\istate}{s}
\newcommand{\istates}{\Sigma}
\newcommand{\rstate}{\istate^{\sharp}}
\newcommand{\rstates}{\istates^{\sharp}}

\newcommand{\ASSIGN}[2]{{#1} \leftarrow {#2}}
\newcommand{\TASSIGN}[2]{{#1} \leftarrow {#2}}
\newcommand{\HAVOC}[1]{\ASSIGN{#1}{\NONDET()}}
\def\ASSUME{{{\mathtt{assume}}}}
\def\ASSERT{{{\mathtt{assert}}}}
\newcommand{\GET}[2]{\TASSIGN{#2}{#1}}
\newcommand{\SET}[2]{\TASSIGN{#1}{#2}}

\def\SKIP{{{\mathtt{skip}}}}
\def\LOOP{{{\mathtt{loop}}}}

\def\vals{V}
\def\val{v}

\def\CHOOSE{{[\!]}}
\def\instr{\mathtt{I}}
\def\prgm{\mathtt{P}}
\def\rkvar{{\kvar}}

\def\RELSEM{{Relational}\xspace}
\def\IMPSEM{{Imperative}\xspace}
\def\REFREL{relation\xspace}

\def\ERROR{{\mathcal{E}}}
\def\CHOICE{{\mathrm{[\!] }}}
\def\EXPAND{{\sf Expand}}
\def\CLONE{{\sf Clone}}
\def\rpost{\mathsf{Post}^{\sharp}}
\def\rreach{\mathsf{Reach}^{\sharp}}
\def\ipost{\mathsf{Post}}
\def\ireach{\mathsf{Reach}}

\begin{document}

\title{Refinement Type Inference via Abstract Interpretation}

\authorinfo{Ranjit Jhala}{UCSD}{jhala@cs.ucsd.edu}
\authorinfo{Rupak Majumdar}{UCLA}{rupak@cs.ucla.edu}
\authorinfo{Andrey Rybalchenko}{TUM}{rybal@in.tum.de}
\maketitle

\begin{abstract}
Refinement Types are a promising approach for checking behavioral
properties of programs written using advanced language features like
higher-order functions, parametric polymorphism and recursive datatypes.
The main limitation of refinement type systems to date is the  
requirement that the programmer provides the types of all functions, 
after which the type system can {\em check} the types and hence, verify the program.

In this paper, we show how to automatically {\em infer} refinement types, using
existing abstract interpretation tools for imperative programs. 
In particular, we demonstrate that the problem of refinement type 
inference can be reduced to that of computing invariants of simple, 
first-order imperative programs without recursive datatypes.
As a result, our reduction shows that any of the wide variety of 
abstract interpretation techniques developed for imperative programs, 
such as polyhedra, counterexample guided predicate abstraction 
and refinement, or Craig interpolation,
can be directly applied to verify behavioral properties of 
modern software in a fully automatic manner.
\end{abstract}

\section{Introduction}
\label{sec:intro}

Automatic verification of semantic properties of modern programming languages
is an important step toward reliable software systems.
For higher-order programming languages with inductive datatypes
or polymorphic instantiation, the main verification tool has been type systems,
which traditionally capture only coarse data-type properties (such as $\ttint$s are
only added to $\ttint$s),
and require the programmer to explicitly annotate program invariants if
more precise invariants about program computations are required. 

For example, \emph{refinement} type systems \cite{XiPfenning99} 
associate data types with refinement predicates that capture richer properties of
program computation.
Using refinement types, one can state, for instance, that a program variable $\ttxs$ has the refinement type
``non-zero integer," or that the integer division function has the refinement type 
$\ttint \rightarrow \reftyp{\valu}{\ttint}{\valu \not = 0} \rightarrow \ttint$ 
which states that the second argument must be non-zero.
Then if a program with refinement type type-checks, one can assert that there is no
division-by-zero error in the program.
The idea of refinement types to express precise program invariants is
well-known~\cite{XiPfenning99,Ou2004,ATS,Dunfield,Flanagan06,GordonRefinement09}.
However, in each of the above systems, the programmer must provide refinements for
each program type, and the type system {\em checks} the provided type refinements for
consistency.
We believe that this burden of annotations has limited the widespread adoption of refinement
type systems.

For {\em imperative} programming languages, algorithms based on abstract interpretation
can be used to {\em automatically infer} many program invariants 
\cite{SLAMPOPL02,HJMM04,CousotPLDI03}, thereby proving many semantic properties of practical interest.
However, these tools do not precisely model modern programming features such as closures
and higher-order functions or inductive datatypes, and so in practice, they
are too imprecise when applied to higher-order programs.

In this paper, we present an algorithm to {\em automatically} 
verify properties of higher-order programs through
refinement type inference (RTI) by
combining refinement type systems for higher-order programs
with invariant synthesis techniques for first-order programs.
Our main technical contribution is a translation 
from type constraints derived from a refinement type system for
higher-order programs to a first-order imperative program with assertions,
such that the assertions hold in the first-order program
iff there is a refinement type that makes the higher-order program
type-check.
Moreover, a suitable type refinement for the higher-order program
can be constructed from the invariants of the first-order program.
Thus, our algorithm replaces the manual annotation burden for refinement types with
automatically constructed program invariants on the translated program,
thus enabling fully automatic verification of programs written 
in modern languages.

\begin{figure}[t]
  \vspace{1ex}
  \centering
\begin{minipage}[t]{.8\columnwidth}
\tikzset{
    state/.style={
           rectangle,
           rounded corners,
           draw=black, thick, 
           minimum height=2em,
           minimum width=10em,
           inner sep=2pt,
           text centered,
           },
}
\begin{tikzpicture}[->,>=stealth',shorten >=1pt,auto, node distance=1.2cm,
                    semithick]
  \tikzstyle{every state}=[draw=black]

  \node[draw=none] (ML) {
    \begin{minipage}[t]{.9\columnwidth}
      \centering
      OCaml Program\\(with assertions)
    \end{minipage}
  };
  \node[state,below of=ML] (Gen) {Constraint Generation};
  \path (ML) edge (Gen);
  \node[state,below of=Gen,double] (RTI) {RTI Translation};
  \path (Gen) edge  node[right]{Subtyping Constraints} (RTI);
  \node[state,below of=RTI] (AI) {Abs. Interpretation};
  \path (RTI) edge node[right]{Simple IMP Program} (AI);
  \node[draw=none,node distance=1.5cm,below left of=AI] (Safe) {Safe};
  \path (AI) edge  (Safe);
  \node[draw=none,node distance=1.5cm,below right of=AI] (Unsafe) {Unsafe};
  \path (AI) edge  (Unsafe);

  \node at (RTI.center) {
    \fbox{
    \begin{minipage}[t]{19em}
      \mbox{}
      \vspace{10.5em}
    \end{minipage}}
  };

\end{tikzpicture}
\end{minipage}

\caption{RTI algorithm.}
\label{fig:algo}
\end{figure}
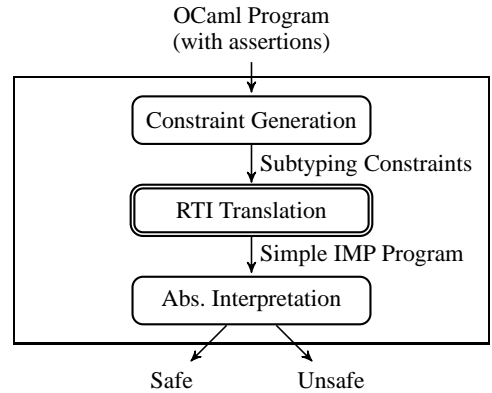

\noindent 
The RTI algorithm (Figure~\ref{fig:algo}) 
proceeds in three steps.

\mypara{Step 1: Type-Constraint Generation.} 
First, it performs Hindley-Milner type inference \cite{Milner82} to construct
\ML types for the program, and uses these types to generate 
\emph{refinement templates}, \ie types in which 
\emph{refinement variables} $\kvar$ are used 
to represent the unknown refinement predicates.
Then, the algorithm uses a standard syntax-directed procedure 
to generate subtyping constraints over the templates
such that the program type checks (\ie is safe) if
the subtyping constraints are satisfiable
\cite{XiPfenning99,Knowles07,LiquidPLDI08,GordonRefinement09}.

\mypara{Step 2: Translation.}
Second, it translates the set of type constraints 
to a \emph{first-order, imperative program over base values}
such that the type constraints are satisfiable if and only if 
the imperative program does not violate any assertions.

\mypara{Step 3: Abstract Interpretation.}
Finally, an abstract interpretation technique for first order
imperative programs is used to prove that the first order 
program is safe.
The proof of safety produced by this analysis automatically
translates to solutions to the
refinement type variables, thus generating refinement types for the
original \ML program.

The main contribution of this paper is the RTI translation algorithm.
The advantage of the translation is that 
it allows one to apply any of the well-developed semantic 
imperative program analyses based on abstract interpretation (\eg 
polyhedra~\cite{CousotHalbwachs78} and octagons~\cite{CousotPLDI03}, 
counterexample-guided predicate abstraction refinement
(CEGAR)~\cite{SLAMPOPL02,HJMM04}, 
Craig interpolation~\cite{HJMM04,McMillan06}, 
constraint-based invariant
generation~\cite{Sankaranarayanan05,RybalchenkoVMCAI07}
random interpretation~\cite{Gulwani03}, 
\etc) 
to the verification of modern software with 
polymorphism, inductive datatypes, and higher-order functions.
Instead of painstakingly reworking each semantic analysis 
for imperative programs to the higher order setting,
possibly re-implementing them in the process, one 
can use our translation, and apply any existing analysis as is. 
In fact, using the translation, our implementation {\em directly} 
uses a CEGAR and interpolation based safety verification tool 
to verify properties of \ocaml programs.

In essence, our algorithm separates syntactic reasoning about function calls
and inductive data types (handled well by typing constraints) from
semantic reasoning about data invariants (handled well by abstract domains). 
The translation from refinement type constraints to 
imperative programs in Step~2 is the key enabler. 
The translation, and the proof that the satisfiability of type constraints and 
safety of the translated program are equivalent, are based on the following 
observations.

The first observation is that refinement type 
variables $\kvar$ define \emph{relations} over the value being 
defined by the refinement type
and the finitely many variables that are in-scope at the 
point where the type is defined.
In the imperative program, each finite-arity relation can be encoded 
with a variable that encodes a relation.
Each refinement type constraint can be encoded as a straight-line 
sequence that reads tuples from and writes tuples to the relation variables,
and the set of constraints can be encoded as a non-terminating while-loop
that in each iteration, non-deterministically executes 
one of the blocks.
Thus, the problem of determining the existence of appropriate relations
reduces to that of computing (overapproximations) of the set of tuples
in each relation variable in the translated
program~(Theorem~\ref{th:translate}). 

Our second observation is that if the translated program is in a special 
\emph{read-write-once} form, where within each straight-line block
a relation variable is read and written \emph{at most once}, 
then one can replace all relation-valued variables with variables whose
values range over tuples~(Theorem~\ref{th:rwo-equiv}). 
Moreover, we prove that we can, without affecting satisfiability, 
preprocess the refinement typing constraints so that the translated program is 
a read-write-once program~(Theorem~\ref{th:clone}).
Together, the observations yield a simple and direct translation
from refinement type inference to simple imperative programs.

We have instantiated our algorithm in a verification tool for \ocaml programs.
Our implementation 
generates refinement type constraints using the algorithm of~\cite{LiquidPLDI08}, 
and uses the \ARMC~\cite{PADL07} software model checker to verify the translated programs. 
This allows fully automatic verification of a set of \ocaml benchmarks
for which previous approaches either required manual annotations
(either the refinement types \cite{XiPfenning99} or their constituent
predicates \cite{LiquidPLDI08}), or an elaborate customization and
adaptation of the counterexample-guided abstraction refinement
paradigm~\cite{TerauchiPOPL2010}.
Thus, we show, for the first time, how abstract interpretation can be
lifted ``as-is'' to the practical refinement type inference for
modern, higher-order languages.

While we have focused on the verification of 
functional programs, our approach is language independent, 
and requires only an appropriate refinement type system for the source
language. 

\section{Overview}

\begin{figure}[t]
\begin{small}
\begin{center}
\begin{verbatim}
    let rec iteri i xs f = 
      match xs with
      | []     -> ()
      | x::xs' -> f i x; 
                  iteri (i+1) xs' f

    let mask a xs = 
      let g j y = a.(j) <- y && a.(j) in
      if Array.length a = List.length xs then
        iteri 0 xs g
\end{verbatim}
\end{center}
\end{small}
\caption{\ML Example}
\label{ex:ml-abc}
\end{figure}

We begin with an example that illustrates how our refinement
type inference (\HMC) algorithm combines type
constraints and abstract interpretation to automatically verify safety properties of
\emph{functional} \ML programs with higher-order functions and 
recursive structures.
We show that the combination of syntactic type constraints
and semantic abstract interpretation enables the automatic verification of properties
that are currently beyond the scope of either technique in isolation.

\mypara{An \ML Example. }
Figure~\ref{ex:ml-abc}(a) shows a simple ML program that 
updates an array $\tta$ using the elements of the list $\ttxs$. 
The program comprises two functions. 
The first is a higher-order list \emph{indexed-iterator}, $\ttiteri$,
that takes as arguments a starting index $\tti$, 
a (polymorphic) list $\ttxs$, 
and an iteration function $\ttf$. 
The iterator goes over the elements of the list and invokes $\ttf$ on each element
and the index corresponding to the element's position in the list.
The second is a client, $\ttmask$, of the iterator $\ttiteri$ that takes as input a
boolean array $\tta$ and a list of boolean values $\ttxs$, and if the
lengths match, calls the indexed iterator with an iteration function $\ttg$ 
that masks the $\ttj^{th}$ element of the array.

Suppose that we wish to statically verify the safety of the array reads and writes 
in function $\ttg$; that is to prove that whenever $\ttg$ is invoked, 
$0 \leq \ttj < \ttlena(\tta)$.
As this example combines higher-order functions, recursion, data-structures, and
arithmetic constraints on array indices, it is difficult to analyze automatically 
using either existing type systems or abstract interpretation implementations in isolation.
The former do not precisely handle arithmetic on indices, and the latter
do not precisely handle higher-order functions and are often imprecise on
data structures.
We show how our \HMC technique can automatically prove the correctness 
of this program.

\mypara{Refinement Types.} 
To verify the program, we compute program invariants that are expressed 
as \emph{refinements} of \ML types with predicates over program 
values \cite{Knowles07,GordonRefinement09,LiquidPLDI08}.
The predicates are additional constraints that must be satisfied by 
every value of the type. A base value, say of type $\ttint$, 
can be described by the refinement type
$\reftyp{\valu}{\ttint}{p}$ 
where $\valu$ is a special \emph{value variable} representing the type
being defined, and $p$ is a \emph{refinement predicate} which constrains 
the range of $\valu$ to a subset of integers.
For example, the type 
$\reftyp{\valu}{\ttint}{0 \leq \valu < \ttlena(\tta)}$ 
denotes the set of integers $c$ that are between $0$ and the 
value of the expression $\ttlena(\tta)$.
Thus, the unrefined type $\ttint$ abbreviates $\reftyp{\valu}{\ttint}{\true}$, 
which does not constrain the set of integers.
Base types can be combined to construct \emph{dependent function types}, 
written $\ftyp{\ttx}{\tliqs_1} \rightarrow \tliqs_2$,
where $\tliqs_1$ is the type of the domain, 
$\tliqs_2$ the type of the range, 
and where the name $\ttx$ for the 
formal parameter can appear in the refinement predicates 
in $\tliqs_2$.
For example, the type 
$$\ftyp{\ttx}{\reftyp{\valu}{\ttint}{\valu\geq 0}} \rightarrow \reftyp{\valu}{\ttint}{\valu = \ttx+1}$$ 
is the type of a function which takes a non-negative integer parameter and returns an 
output which is one more than the input.
In the following, we write $\typ$ for the type $\reftyp{\valu}{\typ}{\true}$.
When $\valu$ and $\typ$ are clear from the context, we write
$\sreftyp{p}$ for $\reftyp{\valu}{\typ}{p}$.

\mypara{Safety Specification.}
Refinement types can be used to \emph{specify} safety properties by 
encoding pre-conditions into primitive operations of the language.
For example, consider the array read $\mathtt{\tta.(\ttj)}$ (resp.\
write $\mathtt{\tta.(\ttj) \leftarrow \tte}$) in $\ttg$ which is an
abbreviation for ${\ttget\ \tta\ \ttj}$ (resp.\ ${\ttset\ \tta\ \ttj\
\tte}$).
By giving $\ttget$ and $\ttset$ the refinement types
\begin{align*}
& \ftyp{\tta}{\alpha \ttarray} \rightarrow 
{\reftyp{\valu}{\ttint}{0 \leq \valu < \ttlena(\tta)}} \rightarrow 
\alpha\ , \\
& \ftyp{\tta}{\alpha \ttarray} \rightarrow
{\reftyp{\valu}{\ttint}{0 \leq \valu < \ttlena(\tta)}} \rightarrow
\alpha \rightarrow 
\ttunit\ ,
\end{align*}
we can specify that in any program the array accesses must be within
bounds. More generally, arbitrary safety properties can be specified 
by giving $\ttassert$ the appropriate refinement type \cite{LiquidPLDI08}.

\mypara{Safety Verification.}
The \ML type system is too imprecise to prove the safety 
of the array accesses in our example as it infers that $\ttg$ 
has type
${\ftyp{\ttj}{\ttint} \rightarrow \ftyp{\tty}{\ttbool} \rightarrow \ttunit}$,
\ie that $\ttg$ can be called with \emph{any} integer $\ttj$.
If the programmer manually provides the refinement types
for all functions and polymorphic type instantiations,
refinement-type checking~\cite{XiPfenning99,Dunfield,GordonRefinement09}
can be used to verify that the provided types were consistent
and strong enough to prove safety. 
This is analogous to providing pre- and post-conditions and loop-invariants 
for verifying imperative programs.
For our example, the refinement type system could check the program 
if the programmer provided the types:
$$
\begin{array}{rl}
\ttiteri \hastype & \ftyp{\tti}{\ttint} \rightarrow
	 	      \ftyp{\ttxs}{\reftyp{\valu}{\alpha\ \ttlist}{0 \leq \ttlenl(\valu)}} \rightarrow \\
         	    & (\ftyp{\ttj}{\sreftyp{\tti \leq \valu < \ttlenl(\ttxs)}} \rightarrow \alpha \rightarrow \ttunit) \rightarrow \ttunit\\
\ttg \hastype & \ftyp{\ttj}{\sreftyp{0 \leq \valu < \ttlena(\tta)}} \rightarrow \ttbool \rightarrow \ttunit\\
\end{array}
$$
Here, we omitted refinement predicates that are equal to true, e.g.,
for \texttt{i} in the type of~\texttt{iteri}.

\mypara{Automatic Verification via \HMC.}
As even this simple example illustrates, the type annotation burden for
verification is extremely high.
Instead, we would like to verify the program without requiring the programmer
to provide every refinement type.
%
The \HMC algorithm proceeds in three steps. 
First, we syntactically analyze the \emph{source} program to generate 
subtyping constraints over refinement templates.
Second, we translate the constraints into an equivalent simple imperative
\emph{target} program.
Third, we semantically analyze the target program to determine whether
it is safe, from which we conclude that the constraints are satisfiable and
hence, the source program is safe.
Next, we illustrate these steps using Figure~\ref{ex:ml-abc} as
the source program.

\subsection{Step 1: Constraint Generation}
In the first step, we generate a system of refinement type constraints 
for the source program \cite{Knowles07,LiquidPLDI08}.
To do so, we
(a)~build templates that refine the \ML types with 
refinement variables that stand for the unknown refinements, and
(b)~make a syntax-directed pass over the program to generate subtyping
constraints that capture the flow of values. 
For the functions $\ttiteri$ and $\ttg$ from Figure~\ref{ex:ml-abc}, 
with the respective \ML types
\begin{align*}
&\ftyp{\tti}{\ttint} 
 \rightarrow \ftyp{\ttxs}{\alpha\ \ttlist} 
 \rightarrow (\ftyp{\ttj}{\ttint} \rightarrow \alpha \rightarrow \ttunit) \rightarrow
 \ttunit \\
& \ftyp{\ttj}{\ttint} \rightarrow \ttbool \rightarrow \ttunit
\intertext{we would generate the respective templates}
& \ftyp{\tti}{\ttint} 
  \rightarrow \ftyp{\ttxs}{\sreftyp{0 \leq \ttlenl(\valu)}} 
  \rightarrow (\ftyp{\ttj}{\sreftyp{\kvar_1}} \rightarrow \alpha \rightarrow \ttunit) \rightarrow
  \ttunit \\
& \ftyp{\ttj}{\sreftyp{\kvar_2}} \rightarrow \ttbool \rightarrow \ttunit
\end{align*}
Notice that these templates simply refine the \ML types with refinement
variables $\kvar_1, \kvar_2$ that stand for the unknown refinements.
For clarity of exposition, we have added the refinement $\true$ 
for some variables (\eg for the type $\alpha$ and $\ttbool$); 
our system would automatically infer the unknown refinements.
We model the length of lists (resp.\ arrays) with an uninterpreted 
function $\ttlen$ from the lists (resp.\ arrays) to integers, and 
(again, for brevity) add the refinement stating $\ttxs$ 
has a non-negative length in the type of $\ttiteri$. 

After creating the templates, we make a syntax-directed pass over the 
program to generate constraints that capture relationships 
between refinement variables. There are two kinds of type constraints --
{\em well-formedness} and {\em subtyping}.

\mypara{Well-formedness Constraints } 
capture scoping rules, and ensure that the
refinement predicate for a type can only refer to variables that are in scope.
Our example has two constraints:
\begin{align}
\ftyp{\tti}{\ttint}; \ftyp{\ttxs}{\alpha\ \ttlist} \deriv & \reftyp{\valu}{\ttint}{\kvar_1} \label{eq:w1} \tag{w1} \\
\ftyp{\tta}{\ttbool\ \ttarray}; \ftyp{\ttxs}{\alpha\ \ttlist} \deriv & \reftyp{\valu}{\ttint}{\kvar_2} \label{eq:w2} \tag{w2} 
\end{align}
The first constraint states that $\kvar_1$, which represents the unknown 
refinement for the first parameter passsed to the higher-order
iterator $\ttiteri$, can only refer to the two program variables that are in-scope at that
point, namely $\tti$ and $\ttxs$.
Similarly, the second constraint states that $\kvar_2$, which refines 
the first argument of $\ttg$, can only refer to $\tta$ and $\ttxs$, which
are in scope where $\ttg$ is defined.


\mypara{Subtyping Constraints }
reduce the flow of values within the program into subtyping
relationships that must hold between the source and target of the flow.
Each constraint is of the form 
\begin{align}
\renv \deriv & \tliqs_1 \subt \tliqs_2 \notag 
\intertext{where $\renv$ is an \emph{environment} comprising a sequence of type bindings,
and $\tliqs_1$ and $\tliqs_2$ are refinement templates.
The constraint intuitively states that under the environment $\renv$, the
type $\tliqs_1$ must be a subtype of $\tliqs_2$.
The subtyping constraints are generated syntactically from the code.
First consider the function $\ttiteri$.
The call to $\ttf$ generates}
\renv \deriv & \sreftyp{\valu = \tti} \subt \set{\kvar_1} \label{eq:c1} \tag{c1} 
\intertext{where the environment $\renv$ comprises the bindings}
\renv \defeq & \ftyp{\tti}{\sreftyp{\true}};\ \ftyp{\ttxs}{\sreftyp{0 \leq \ttlenl(\valu)}}; \notag \\ 
	     & \ftyp{\ttx}{\sreftyp{\true}};\ \ftyp{\ttxsp}{\sreftyp{0 \leq \ttlenl(\valu) = \ttlenl(\ttxs) - 1}}
  \notag
\intertext{the constraint ensures that at the callsite, 
the type of the actual is a subtype of the formal.
The bindings in the environment
are simply the refinement templates for the variables in scope at the point
the value flow occurs. The type system yields the information 
that the length of $\ttxsp$ is one less than $\ttxs$ as the former is the
tail of the latter \cite{XiPfenning99,LiquidPLDI09}.
Similarly, the recursive call to $\ttiteri$ generates}
\renv \deriv & \set{\ftyp{\ttj:\kvar_1} \rightarrow \alpha \rightarrow \ttunit} \subt \notag \\	
	     & \set{(\ftyp{\ttj}{\kvar_1} \rightarrow \alpha \rightarrow
	     \ttunit)[\tti+1/\tti][\ttxsp/\ttxs]}
\notag
\intertext{which states that type of the actual $\ttf$ is 
a subtype of the third formal parameter of $\ttiteri$
after applying substitutions
$\SUBST{}{\tti}{\tti+1}$ and $\SUBST{}{\ttxs}{\ttxsp}$
that capture the passing in of the actuals 
$\tti+1$ and $\ttxsp$ for the first two parameters respectively.
By pushing the substitutions inside and applying the standard rules for 
function subtyping, this constraint simplifies to}
\renv \deriv & \set{\SUBST{\SUBST{\kvar_1}{\tti+1}{\tti}}{\ttxsp}{\ttxs}} \subt
\set{\kvar_1} \label{eq:c2} \tag{c2}
\end{align}
Next, consider the function $\ttmask$. The array accesses inside $\ttg$ 
generate the ``bounds-check" constraint
\begin{align}
\renv'; \ftyp{\ttj}{\set{\kvar_2}}; \ftyp{\tty}{\sreftyp{\true}} \deriv & \sreftyp{\valu = \ttj} \subt 
\sreftyp{0 \leq \valu < \ttlena(\tta)} \label{eq:c3} \tag{c3} 
\intertext{where $\renv' \defeq \ftyp{\tta}{\ttbool\ \ttarray}; \ftyp{\ttxs}{\sreftyp{0 \leq \ttlenl (\valu)}}$ 
has bindings for the other variables in scope. 
Finally, the flow due to the third parameter for the call to $\ttiteri$ yields}
\renv'; \ttlen(\tta) = \ttlenl(\ttxs) \deriv 
& \set{\ftyp{\ttj}{\kvar_2} \rightarrow \typ} \subt \notag \set{\SUBST{(\ftyp{\ttj}{\kvar_1} \rightarrow \typ)}{\tti}{0}} 
\intertext{where for brevity we write $\typ$ for $\ttbool \rightarrow
\ttunit$, and omit the trivial substitution $\SUBST{}{\ttxs}{\ttxs}$ due to
the second parameter.
The last conjunct in the environment captures the guard
from the $\ttif$ under whose auspices the call occurs. 
By pushing the substitutions inside and 
applying standard function subtyping, the above reduces to}
\renv'; \ttlen(\tta) = \ttlenl(\ttxs) \deriv 
& \set{\SUBST{\kvar_1}{\tti}{0}} \subt \set{\kvar_2}
\label{eq:c4} \tag{c4} 
\end{align}
For brevity we omit trivial constraints like ${\cdot \deriv \ttint \subt \ttint}$.
If the set of constraints constructed above is satisfiable,
then there is a valid refinement typing of the program~\cite{LiquidPLDI08},
and hence the program is safe.

\subsection{Step 2: Translation to Imperative Program}

Determining the satisfiability of the constraints 
requires semantic analysis about program computations. 
In the second step, our key technical contribution, 
we show a translation that reduces the constraint 
satisfiability problem to checking the safety of 
a simple, imperative program. 
Our translation is based on two observations.

\mypara{Refinements are Relations.}
The first observation is that type refinements are defined through
{\em relations}: 
the set of values denoted by a refinement type 
$\reftyp{\valu}{\typ}{p}$ where $p$ refers to the program 
variables $\ttx_1,\ldots,\ttx_n$ 
of the respective base types $\typ_1,\ldots,\typ_n$ 
is equivalent to the set 
$$
\set{t_0 \mid \exists (t_1,\ldots,t_n) \mbox{ s.t.\ }
  \begin{array}[t]{@{}l}
    (t_0, t_1,\ldots,t_n) \in R_p \mathrel{\wedge} \\
    \quad t_1 = \ttx_1 \wedge \ldots t_n = \ttx_n
  \end{array}
}
$$
where $R_p$ is an $(n+1)$-ary relation in $\typ \times \typ_1\times\ldots\times \typ_n$ defined by 
$p$.
For example, the set of values denoted by 
$\reftyp{\valu}{\ttint}{\valu \leq \tti}$
is equivalent to the set:
$$\set{t_0 \mid \exists t_1 \mbox{ s.t.\ } (t_0, t_1) \in R_{\leq} \wedge t_1 = \tti}\ ,$$
where $R_\leq$ is the standard $\leq$-ordering relation over the integers.
In other words, each refinement variable $\kvar$ can be seen as 
the projection on the first co-ordinate of
a $(n+1)$-relation over the variables $(\valu, x_1,\ldots,x_n)$, 
where $x_1,\ldots,x_n$ are the variables in the well-formedness 
constraint for $\kvar$ (\ie the variables in scope of $\kvar$). 
Thus, the problem of determining the satisfiability of the constraints
is analogous to the problem of determining the existence of appropriate 
relations.

\mypara{Relations are Records.}
The second observation is that the problem of finding appropriate relations
can be reduced to the problem of analyzing a simple imperative program 
with variables ranging over relations.
In the imperative program, each refinement variable,
standing for an $n$-ary relation, is translated into a record variable with
$n$-fields. 
Each subtyping constraint can be translated into a block of reads-from 
and writes-to the corresponding records.
The set of all tuples that can be written into a 
given record on some execution of the program defines the corresponding relation. 
The entire program is an infinite loop, which in each iteration
non-deterministically chooses a block of reads and writes defined by 
a constraint.

The arity of a relation,
and hence the number of fields of the corresponding record, is determined by
the well-formedness constraints. 
For example, the constraint~\eqref{eq:w1} specifies that $\kvar_1$
corresponds to a ternary relation, that is, a set of triples 
where the $0^{th}$ element (corresponding to $\valu$) is an integer, 
the $1^{st}$ element (corresponding to $\tti$) is an integer, 
and the $2^{nd}$ element (corresponding to $\ttxs$) is a list.
We encode this in the imperative program via a record variable $\rkvar_1$
with three fields $\rkvar_1.0$, $\rkvar_1.1$ and $\rkvar_1.2$.

Figure~\ref{ex-ml-imperative} shows the imperative program translated from
the constraints for our running example. 
We use the subtyping constraints to define the flow of tuples into records.
For example, consider the constraint~\eqref{eq:c2} which is translated
to the block marked \verb+/*c2*/+.
Each variable in the type environment is translated to a corresponding
variable in the program. 
The block has a sequence of assignments that define the environment
variables.
For example, we know $\tti$ has type $\ttint$, 
so there is an assignment of an arbitrary integer to $\tti$.
When there is a known refinement in the binding, the non-deterministic assignment is followed by an
assume operation (a conditional) that establishes that the value
assigned satisfied the given refinement.
For example $\ttxs$ gets assigned an arbitrary value, but then the assume
establishes the fact that the length of $\ttxs$ is non-negative.
Similarly $\ttxsp$ gets assigned an arbitrary value, that has non-negative
length and whose length is 1 less than that of $\ttxs$.
The LHS of \eqref{eq:c2} reads a tuple from $\rkvar_1$ whose first and
second fields are assumed to equal the ${\tti+1}$ and $\ttxsp$
respectively.
Finally, the triple $(\valu, \tti, \ttxs)$ is written into the record
$\rkvar_1$ which is the RHS of \eqref{eq:c2}.

Next, consider the translated block for the bounds-check constraint
\eqref{eq:c3}. Here, the translation is as before but the
RHS is a known refinement predicate (that stipulates the integer be within
bounds). In this case, instead of writing into the record that defines the
RHS, the translation contains an assertion over the corresponding variables
that ensures that the refinement predicate holds. 

\mypara{\RELSEM vs. \IMPSEM Semantics.}
There is a direct correspondence between the
refinement-relations and the record variables when the 
translated program is interpreted under a \RELSEM semantics,
where 
(1)~the records range over (initially empty) \emph{sets of tuples},
(2)~each write adds a new tuple to the record's set, and, 
(3)~each read non-deterministically selects some tuple from the record's set.
Under these semantics, we can show that the constraints 
are satisfiable iff the imperative program is safe (\ie no assert fails on any execution)
(Theorem~\ref{th:translate}).

Unfortunately, these semantics preclude the direct application of mature
invariant generation and safety verification techniques 
\eg those based on abstract interpretation or CEGAR-based 
software model checking, as those techniques 
do not deal well with set-valued variables.
We would like to have an imperative semantics where each record 
contains a single value, the last tuple written to it.
We show that there is a syntactic subclass of programs for which
the two semantics coincide. 
That is, a program in the subclass is safe under the imperative
semantics if and only if it is safe under the set-based semantics
(Theorem~\ref{th:rwo-equiv}).
Furthermore, we show a technique that ensures that the translated program
belongs to the subclass (Theorem~\ref{th:clone}).

The attractiveness of the translation is that the resulting 
programs fall in a particularly pleasant subclass of programs 
which do not have any advanced language features like
higher-order functions, polymorphism, and recursive data structures,
or variables over complex types such as sets,
that are the bane of semantic analyses.
Thus, the translation yields simple imperative programs to which 
a wide variety of semantic analyses directly apply.

\subsection{Step 3: Invariant Generation.}
Together these results imply that we can run off-the-shelf 
abstract interpretation and invariant generation tools on the 
translated program, and use the result of the analysis to 
determine whether the original \ML program is typable. 

For the translated program shown in Figure~\ref{ex-ml-imperative}, 
the CEGAR-based software model checker \ARMC~\cite{PADL07} 
finds that the assertion is never violated, and
computes the invariants:
\begin{align*}
& \rkvar_1.1 \leq \rkvar_1.0 \wedge \rkvar_1.0 < \ttlen(\rkvar_1.2)  \\
& 0 \leq \kvar_2.0 < \ttlen (\rkvar_2.1)
\end{align*}
which, when plugging in $\valu$, $\tti$ and $\ttxs$ for the 
$0^{th},1^{st},2^{nd}$ fields of $\rkvar_1$ 
and $\valu$, $\tta$ for the $0^{th}, 1^{st}$ fields of $\rkvar_2$
respectively, yields the refinements
\[
\kvar_1 \defeq  \tti \leq \valu < \ttlen(\ttxs) \ \ \ 
\kvar_2 \defeq  0 \leq \valu < \ttlen (\tta)
\]
which suffice to typecheck the original \ML. 
Indeed, these predicates for $\kvar_1$ and $\kvar_2$ are easily shown to satisfy the
constraints (c1), (c2), (c3), and (c4).

\begin{comment}
\mypara{Outline.}
This concludes a high-level overview of the \HMC approach. 
In the rest of the paper, we start by formalizing the 
notions of refinement predicates, type constraints and constraint
satisfaction (Section~\ref{sec:constraints}.
Next, we describe the syntax and semantics of the target imperative
programs (Section~\ref{sec:imp}).
Next, we make precise our translation, and prove the equivalence of
the (source) type constraints and the (target) translated program
(Section~\ref{sec:equiv}).
After that, we describe our prototype implementation and preliminary
experiments using \HMC to verify small but complex examples
(Section~\ref{sec:experiments}), and we conclude with a discussion of
the ramifications of our technique and connections to related work
(Section~\ref{sec:discussion}).
\end{comment}

\begin{figure}[t]
\begin{small}
\[
\begin{array}{rl}
\LOOP\{ & \mathtt{{/*} c1 {*/}} \\
	& \HAVOC{\tti};\\
        & \HAVOC{\ttxs};\ \ASSUME(0 \leq \ttlen(\ttxs));\\
	& \HAVOC{\ttxsp};\ \ASSUME(0 \leq \ttlen(\ttxsp) = \ttlen(\ttxs)-1);\\
	& \HAVOC{\valu};\ \ASSUME(\valu = \tti);\\
	& \SET{\rkvar_1}{(\valu,\tti,\ttxs)}\\[4pt]
\CHOICE & \mathtt{{/*} c2 {*/}} \\
	& \HAVOC{\tti};\\
        & \HAVOC{\ttxs};\ \ASSUME(0 \leq \ttlen(\ttxs));\\
	& \HAVOC{\ttxsp};\ \ASSUME(0 \leq \ttlen(\ttxsp) = \ttlen(\ttxs)-1);\\
	& \GET{\rkvar_1}{(t_0,t_1,t_2)}; \\
	& \ASSUME(t_1 = \tti +1);\\
	& \ASSUME(t_2 = \ttxsp);\\
	& \ASSIGN{\valu}{t_0};\\
	& \SET{\rkvar_1}{(\valu, \tti, \ttxs)}\\[4pt]
\CHOICE & \mathtt{{/*} c3 {*/}}\\
	& \HAVOC{\tta};\\
        & \HAVOC{\ttxs};\ \ASSUME(0 \leq \ttlen(\ttxs));\\
  	& \GET{\rkvar_2}{(t_0, t_1, t_2)};\\ 
  	& \ASSIGN{\ttj}{t_0};\\
	& \ASSERT(0 \leq j < \ttlen(\tta))\\[4pt]
\CHOICE & \mathtt{{/*} c4 {*/}}\\
	& \HAVOC{\tta};\\
        & \HAVOC{\ttxs};\ \ASSUME(0 \leq \ttlen(\ttxs));\\
	& \ASSUME(\ttlen(\tta) = \ttlen(\ttxs)); \\
	& \GET{\rkvar_1}{(t_0, t_1, t_2)};\\
	& \ASSUME(t_1 = 0);\\
	& \ASSUME(t_2 = \ttxs);\\
	& \ASSIGN{\valu}{t_0};\\
	& \SET{\rkvar_2}{(\valu, \tta, \ttxs)} \\
\}	&
\end{array}
\]
\end{small}
\caption{Translated Program}
\label{ex-ml-imperative}
\end{figure}

\section{Constraints}\label{sec:constraints}

We start by formalizing constraints over types refined with predicates. 
To this end, we make precise the notions of 
refinement predicates (Section~\ref{sec:logic}),
refinement types 
(Section~\ref{sec:reftypes}),
constraints over refinement types 
and the notion of satisfaction 
(Section~\ref{sec:refconstr}).

A discussion of how such constraints can be generated in a syntax-guided
manner from program source
is outside the scope of this paper; we refer the reader to the large body
of prior research that addresses this 
issue~\cite{XiPfenning99,Knowles07,LiquidPLDI08,GordonRefinement09}.

\mypara{Notation.}
We use uppercase ($Z$) to denote sets, lowercase $z$ to denote
elements, and $\mybar{Z}$ for a sequence of elements in~$Z$. 

\subsection{Refinement Logic}
\label{sec:logic}

Figure~\ref{fig:logic} shows the syntax of refinement predicates. 
In our discussion, we restrict the predicate language to the typed quantifier-free
logic of linear integer arithmetic and uninterpreted functions.
However, it is straightforward to extend the logic to include 
other domains equipped with effective decision procedures 
and abstract interpreters.

\mypara{Types and Environments.}
Our logic is equipped with a fixed 
set of \emph{types} denoted $\typ$, 
comprising the basic types
$\ttint$ for \emph{integer} values, 
$\ttbool$ for \emph{boolean} values,
and $\ttunint$, a family of \emph{uninterpreted types} that are used to
encode complex source language types such as products, sums, polymorphic 
type variables, recursive types \etc.
We assume there is a fixed set of uninterpreted functions.
Each uninterpreted function $\ttf$ has a fixed type 
$\typ_\ttf \defeq \mybar{\typ^i_\ttf} \rightarrow \typ^o_\ttf$.
An \emph{environment} is a sequence of variable-type bindings.

\mypara{Expressions and Predicates.}
In our logic, \emph{expressions} $e$ comprise variables, linear arithmetic
(\ie addition and multiplication by constants), and applications of 
uninterpreted functions $\ttf$. 
Note that as is standard in semantic program analyses, complex
operations like division or non-linear multiplication be modelled using 
uninterpreted functions.
Finally, \emph{predicates} comprise atomic comparisons of expressions, or
boolean combinations of sub-predicates.
We write $\true$ (resp. $\false$) as abbreviations for $0=0$ (resp. $0=1$).

\mypara{Well-formedness.}
We say that a predicate $p$ is \emph{well-formed} in an environment
$\tenv$ if every variable appearing in $p$ is bound in $\tenv$ and
$p$ is ``type correct'' in the environment $\tenv$.

\mypara{Validity.}
For each type $\typ$, we write $\univ{\typ}$ to denote the set of
concrete values of~$\typ$.
An \emph{interpretation} $\sigma$ is a map from 
variables $x$ to concrete values, and 
functions $\ttf$ to maps from $\univ{\mybar{\typ^i_\ttf}}$ to $\univ{\typ^o_\ttf}$.
We say that $\sigma$ is \emph{valid} under $\tenv$ if 
for each $\ftyp{x}{\typ} \in \tenv$, we have $\sigma(x) \in \univ{\typ}$.
We say that a predicate $p$ is \emph{valid} in an environment $\tenv$, 
if $\sigma(p)$ evaluates to $\true$ for 
every $\sigma$ valid under $\tenv$. 

\subsection{Refinement Types}
\label{sec:reftypes}

Figure~\ref{fig:constraints} shows the syntax of refinement types and 
environments. 

\mypara{Refinements.}
A \emph{refinement} $r$ is either a predicate $p$ drawn from our logic,
or a \emph{refinement variable with pending substitutions}
$\SUBST{\kvar}{x_1}{y_1}\ldots\SUBST{}{x_n}{y_n}$. 
Intuitively, the former represent \emph{known} refinements (or invariants), 
while the latter represent the \emph{unknown} invariants that hold 
of different program values. 
The notion of pending substitutions~\cite{AbadiCardelliCurienLevy,Knowles07} offers a
flexible way of capturing the value flow that arises in the context of
function parameter passing (in the functional setting), or assignment 
(in the imperative setting), even when the underlying 
invariants are unknown.

\mypara{Refinement Types and Environments.}
A \emph{refinement type} $\reftyp{\valu}{\typ}{r}$ is a triple
consisting of a \emph{value variable} $\valu$ denoting the value being
described by the refinement type, a type $\typ$ describing the
underlying type of the value, and a refinement $r$. 
A \emph{refinement environment} $\renv$ is a sequence of refinement type
bindings.

The value variables are special variables distinct from the program
variables, and can occur inside the refinement predicates.
Thus, intuitively, the refinement type describes the set of
concrete values of the underlying type $\typ$ which additionally 
satisfy the refinement predicate. For example, the refinement type:
	$\reftyp{\valu}{\ttint}{\valu \not = 0}$
describes the set of non-zero integers and,
	$\reftyp{\valu}{\ttint}{\valu = \ttx + \tty}$
describes the set of integers whose value equals 
the sum of the values of the (program) variables $\ttx$ and $\tty$.

Note that path-sensitive branch information can be captured by adding 
suitable bindings to the refinement environment. 
For example, the fact that some expression is only evaluated under 
the if-condition that $\ttx > 100$ can be captured in the
environment via a refinement type binding 
$\ftyp{\ttx_b}{\reftyp{\valu}{\ttbool}{\ttx > 100}}$.

\subsection{Refinement Constraints and Solutions}
\label{sec:refconstr}

Figure~\ref{fig:constraints} shows the syntax of refinement constraints.
Our refinement type system has two kinds of constraints.

\mypara{Subtyping Constraints} are of the form
$${\renv \deriv \reftyp{\valu}{\typ}{r_1} \subt \reftyp{\valu}{\typ}{r_2}}$$
Intuitively, a subtyping constraint states that when the program variables satisfy
the invariants described in $\renv$, the set of values described by
the refinement $r_1$ must be \emph{subsumed by} 
the set of values described by the refinement type $r_2$.

\mypara{Well-formedness Constraints} are of the form 
${\tenv \deriv \reftyp{\valu}{\typ}{r}}$. 
Intuitively, a well-formedness constraints states that the refinement $r$
must be a well-typed predicate in the environment $\renv$ extended with the
binding $\ftyp{\valu}{\typ}$ for the value variable.

\mypara{Embedding.}
To formalize the notions of constraint validity and satisfaction, we embed subtyping
constraints into our logic. We define the function $\embed{\cdot}$ that maps
refinement types, environments and subtyping constraints to predicates in
our logic.
\begin{displaymath}
\begin{array}{ll}
\embed{\reftyp{\valu}{\typ}{p}} 	& \defeq p \\
\embed{\EXT{\ftyp{x}{\tliqs}}{\renv}} 	& \defeq \SUBST{\embed{\tliqs}}{x}{\valu} \wedge \embed{\renv} \\
\embed{\emptyset}			& \defeq \true \\
\embed{\renv \deriv \tliqs_1 \subt \tliqs_2} & \defeq \embed{\renv}
\Rightarrow \embed{\tliqs_1} \Rightarrow \embed{\tliqs_2}\\
\end{array}
\end{displaymath}
Similarly, we define the function $\shape{\cdot}$ that maps refinement
types and environments to types and environments in our logic.
\begin{displaymath}
\begin{array}{ll}
\shape{\reftyp{\valu}{\typ}{p}} 	& \defeq \typ \\
\shape{\EXT{\ftyp{x}{\tliqs}}{\renv}} 	& \defeq \EXT{\ftyp{x}{\shape{\tliqs}}}{\shape{\renv}}\\
\shape{\emptyset}			& \defeq \emptyset \\
\end{array}
\end{displaymath}


\mypara{Validity.}
A subtyping constraint ${\renv \deriv \tliqs_1 \subt \tliqs_2}$ 
that does not contain refinement variables
is \emph{valid} if the predicate 
$\embed{\renv \deriv \tliqs_1 \subt \tliqs_2}$ 
is valid under environment $\shape{\renv}$.
A well-formedness constraint ${\tenv \deriv \reftyp{\valu}{\typ}{p}}$ 
that does not contain refinement variables
is \emph{valid} if the predicate $p$ is well-formed 
in the environment $\tenv$.

\mypara{Relational Interpretations.}
We assume, without loss of generality, that each refinement variable $\kvar$ 
is associated with a unique well-formedness constraint
$\ftyp{x_1}{\typ_1};\ldots;\ftyp{x_n}{\typ_n} \deriv \reftyp{\valu}{\typ_0}{\kvar}$ 
called the well-formedness constraint for $\kvar$.
In this case, we say $\kvar$ has \emph{arity} $n+1$.
Furthermore, we assume that wherever a $\kvar$ of arity $n+1$ appears in 
a subtyping constraint, it appears with a sequence of $n$ pending
substitutions $\SUBST{}{x_1}{y_1} \ldots \SUBST{}{x_n}{y_n}$.
This assumption is without loss of generality, as we can enforce it
with trivial substitutions of the form $\SUBST{}{x_i}{x_i}$.
A \emph{relational interpretation} for $\kvar$ of arity $n+1$, is 
an $(n+1)$-ary relation in $\univ{\typ_0}\times \ldots\times\univ{\typ_n}$.
A \emph{relational model} is a map from refinement variables 
$\kvar$ to relational interpretations.

\mypara{Constraint Satisfaction.}
A set of constraints $C$ is \emph{satisfiable} if 
for all interpretations for uninterpreted functions $\ttf$,
there exists a relational model $\Soln$ such that,
when each occurrence of a refinement type 
$\reftyp{\valu}{\typ}{\SUBST{\kvar}{x_1}{y_1} \ldots \SUBST{}{x_n}{y_n}}$ in $C$ 
is substituted with 
$$\reftyp{\valu}{\typ}{\exists
t_1,\ldots,t_n.\Soln(\kvar)(\valu,t_1,\ldots,t_n) \wedge t_1 = y_1\wedge \ldots t_n = y_n)}$$
every subtyping 
constraint after the substitution is valid. 
In this case, we say that $\Soln$ is a \emph{solution} for $C$.

\begin{figure}[t]
\figbegin
\begin{displaymath}
\begin{array}{lrll}
\typ & ::= &				& \textbf{Types:} \\
  & \mid & \ttint			& \mbox{base type of integers} \\
  & \mid & \ttbool			& \mbox{base type of booleans} \\
  & \mid & \ttunint			& \mbox{complex uninterpreted type} \\[4pt]

\tenv & ::= & 				& \textbf{Environments:} \\
  & \mid & \EXT{\ftyp{x}{\typ}}{\tenv}  & \mbox{binding} \\
  & \mid & \emptyset 			& \mbox{empty} \\[4pt]

e & ::=  &         			& \textbf{Expressions:} \\
  & \mid & x 				& \mbox{variable} \\
  & \mid & n				& \mbox{integer} \\
  & \mid & e_1 + e_2 			& \mbox{addition} \\
  & \mid & n \times e	 		& \mbox{affine multiplication} \\
  & \mid & \ttf(\mybar{e}) 		& \mbox{function application} \\[4pt]

p & ::=  &         			& \textbf{Predicates:} \\
  & \mid & e_1 \rela e_2  		& \mbox{comparison} \\
  & \mid & \neg p 			& \mbox{negation} \\
  & \mid & p_1 \wedge p_2		& \mbox{conjunction} \\
  & \mid & p_1 \Rightarrow p_2		& \mbox{implication} \\[4pt]

r & ::= &				& \textbf{Refinements:} \\
  & \mid & p				& \mbox{predicate} \\
  & \mid & 
  \SUBST{\kvar}{x_1}{y_1}\ldots\SUBST{}{x_n}{y_n} 
  					& \mbox{ref.\ var.\ with substitutions} \\[4pt]
\tliqs & ::= &	\reftyp{\valu}{\typ}{r}	& \textbf{Refinement Types} \\[4pt]
\renv & ::= & 				& \textbf{Refinement Environments:} \\
  & \mid & \EXT{\ftyp{x}{\tliqs}}{\renv}& \mbox{binding} \\
  & \mid & \emptyset 			& \mbox{empty} \\[4pt]
c & ::= & \renv \deriv \tliqs_1 \subt \tliqs_2 	
					& \textbf{Subtype Constraints} \\[4pt]
w & ::= & \tenv \deriv \tliqs	 	& \textbf{WF Constraints}
\end{array} 
\end{displaymath}
\figend
\caption{\textbf{Predicates, Refinements and Constraints.}}
\label{fig:logic}
\label{fig:constraints}
\end{figure}

\section{Imperative Programs}\label{sec:imp}

\HMC translates the satisfiability problem for refinement type constraints
to the question of checking the safety of an imperative program in a simple
imperative language \ilang.
In this section, we formalize the syntax of \ilang programs 
and define the \RELSEM semantics 
and the \IMPSEM semantics. 

%
%

\subsection{Syntax}
\label{sec:impsyntax}

Figure~\ref{fig:impsyntax} shows the syntax of \ilang programs. 
An \emph{instruction} ($\instr$) is a sequence of assignments, assumptions 
and assertions. 
A \emph{program} ($\prgm$) is an infinite loop over a block, whose body
is a non-deterministic choice between a finite number of instructions
$\instr_1,\ldots,\instr_n$.
Next, we describe the different kinds of instructions.
For ease of notation, we assume that there is only one base type $\typ$,
and let $\vals$ denote the set of values of type $\typ$.

\mypara{Variables.} \ilang programs have two kinds of variables. 
(1)~\emph{base} variables, denoted by 
$\valu$, $x$, $y$ and $t$ (and subscripted versions thereof), 
which range over values of type $\typ$.
(2)~\emph{\REFREL} variables, denoted by $\rkvar$,
each of which have a fixed arity $n$ and range over tuples of values
or sets of $n$-tuples of values depending on the semantics.

\mypara{Base Assignments.}
\ilang programs have two kinds of assignments to base variables. 
Either
(1)~an expression over base variables (cf. Figure~\ref{fig:logic}) 
is evaluated and assigned to the base variable, or,
(2)~an arbitrary value of the appropriate base type is assigned to the base
variable, \ie the variable is ``havoc-ed" with a non-deterministically
chosen value.

\mypara{Tuple Assignments.}
The operations \emph{get tuple} and \emph{set tuple} 
respectively read a tuple from and write a tuple to a \REFREL 
variable.

\mypara{Assumes and Asserts.}
\ilang programs have the standard assume and assert instructions using 
predicates over the base variables (cf. Figure~\ref{fig:logic}).
We write $\SKIP$ as an abbreviation for $\ASSUME(0=0)$.

\begin{figure}[t]
\begin{small}
\figbegin
\begin{displaymath}
\begin{array}{llll}
\instr 	& ::=  &  					& \textbf{Instructions:}\\
	& \mid & \ASSIGN{x}{e}				& \mbox{assign expr}\\
	& \mid & \HAVOC{x} 				& \mbox{havoc}\\
	& \mid & \GET{\rkvar}{(t_0,\ldots,t_n)}		& \mbox{get tuple}\\
	& \mid & \SET{\rkvar}{(x_0,\ldots,x_n)}		& \mbox{set tuple}\\ 
	& \mid & \ASSUME(p) 				& \mbox{assume} \\
	& \mid & \ASSERT(p) 				& \mbox{assert} \\
	& \mid & \instr_1; \instr_2			& \mbox{sequence} \\[4pt]
\prgm   & ::=  & \LOOP \{\instr_1 \CHOOSE \ldots \CHOOSE \instr_n \} & \mbox{Program}
\end{array}
\end{displaymath}
\figend
\end{small}
\caption{\textbf{Imperative Programs: Syntax}}
\label{fig:impsyntax}
\end{figure}

\subsection{\RELSEM Semantics}
\label{sec:relsemantics}

We define the \RELSEM semantics as a state transition system.
In this semantics, $\rkvar$ variables  range over 
\emph{sets of} tuples over $\vals$.

\mypara{\RELSEM States.}
A state $\rstate$ in the \RELSEM semantics
is either the special \emph{error state} $\ERROR$ 
or a map from program variables to values such that
every base variable is mapped to a value in $\vals$, 
and every \REFREL variable of arity $n$ is mapped 
to a (possibly empty) set of tuples in $\vals^n$.
Let $\rstates$ be the set of all \RELSEM-program states.

For a state $\rstate$ which is not $\ERROR$, variable $x$ and value $\val$ 
we write $\UPD{\rstate}{x}{\val}$ for the map which maps $x$
to $\val$ and every other key $x'$ to $\rstate(x')$.
We lift maps $\rstate$ from base variables to values to maps 
from expressions (and predicates) to values in in 
the natural way.

\mypara{Initial State.}
The initial state $\rstate_0$ of an \ilang program in the \RELSEM semantics is a map in which
every base variable is mapped to a fixed value from $\vals$,
and every \REFREL variable is mapped to the empty set.

\mypara{Transition Relation.}
The transition relation is defined through a $\rpost$ operator,
shown in Figure~\ref{fig:relsemantics},
which maps a state $\rstate$ and an instruction $\instr$ 
to the \emph{set} of states that the program can be in 
\emph{after} executing the instruction from the state $\rstate$.
We lift $\rpost$ to a set of states $\hat{\rstates}\subseteq \rstates$ in the natural way:
$$\rpost(\hat{\rstates}, \instr) \defeq \bigcup \set{\rpost(\rstate, \instr)\mid \rstate \in \hat{\rstates}}$$
Notice that the program halts if 
a get instruction is executed with an \empty{empty} \REFREL variable, or 
an $\ASSUME(p)$ is executed in a state that does not satisfy $p$.

\mypara{Safety.} 
Let $\prgm$ be the program $\LOOP\{ \instr_1 \CHOOSE \ldots \CHOOSE \instr_n \}$.
The set of \emph{\RELSEM-reachable states} of $\prgm$, denoted $\rreach(\prgm)$ is 
defined by induction as:
$$
\begin{array}{ll}
\rreach(\prgm, 0)   & \defeq  \set{\rstate_0}\\
\rreach(\prgm, m+1) & \defeq  \bigcup \set{\rpost(\rreach(\prgm, m), \instr_j) \mid 1 \leq j \leq n}\\
\rreach(\prgm)      & \defeq  \bigcup \set{\rreach(\prgm, m) \mid 0 \leq m}
\end{array}
$$
A program $\prgm$ is \emph{\RELSEM-safe} if $\ERROR \not \in\rreach(\prgm)$.

\begin{figure*}[t]
\begin{small}
\figbegin
$$\begin{array}{ll}
\multicolumn{2}{l}{\textbf{Common Operations}}\\
\rpost(\ERROR, \instr) 	& \defeq  \set{\ERROR} \\ 
\rpost(\rstate,\instr_1;\instr_2) & \defeq  \rpost(\rpost(\rstate, \instr_1), \instr_2) \\ 
\rpost(\rstate,\ASSIGN{x}{e})  & \defeq  \set{\UPD{\rstate}{x}{\rstate(e)}} \\
\rpost(\rstate,\HAVOC{x}) 	& \defeq  \set{\UPD{\rstate}{x}{c} \mid c \in \vals}\\
\rpost(\rstate,\ASSUME(p)) 	& \defeq  \begin{cases} 
					    	\set{\rstate}   & \mbox{if }\rstate(p) = \true\\
                                       		\emptyset & \mbox{otherwise}
                          	   	  \end{cases}\\
\rpost(\rstate,\ASSERT(p)) 	& \defeq  \begin{cases} 
						\set{\rstate}   & \mbox{if } \rstate(p) = \true\\
                                        	\set{\ERROR} & \mbox{otherwise}
                          	   	  \end{cases}\\[0.2in]
\multicolumn{2}{l}{\textbf{Tuple Operations: \RELSEM Semantics}}\\
\rpost(\rstate,\GET{\rkvar}{(t_0,\ldots,t_n)}) & \defeq
\{\UPD{\rstate}{t_0}{v_0}\ldots\UPD{}{t_n}{v_n} \mid (v_0,\ldots,v_n) \in \rstate(\rkvar)\}\\
\rpost(\rstate,\SET{\rkvar}{(x_0,\ldots,x_n)}) & \defeq 
\set{\UPD{\rstate}{\rkvar}{\rstate(\rkvar) \cup \set{(\rstate(x_0),\ldots,\rstate(x_n))}}}\\[0.2in]
\multicolumn{2}{l}{\textbf{Tuple Operations: \IMPSEM Semantics}}\\
\ipost(\istate,\GET{\rkvar}{(t_0,\ldots,t_n)})	
& \defeq \begin{cases}
	 \set{\UPD{\istate}{t_0}{v_0}\ldots\UPD{}{t_n}{v_n}} & \mbox{if } \istate(\rkvar) = (v_0,\ldots,v_n) \\
	 \emptyset & \mbox{if } \istate(\rkvar) = \bot \\
\end{cases} \\			
\ipost(\istate, \SET{\rkvar}{(x_0,\ldots,x_n)}) & \defeq \set{\UPD{\istate}{\rkvar}{(\istate(x_0),\ldots,\istate(x_n))}}
\end{array}$$
\figend
\end{small}
\caption{\textbf{\RELSEM and \IMPSEM Semantics: Other cases of $\ipost$
identical to $\rpost$}}
\label{fig:impsemantics}
\label{fig:relsemantics}
\end{figure*}

\subsection{\IMPSEM Semantics}
\label{sec:impsemantics}

Next, we define the \IMPSEM semantics, as a state
transition system. In this semantics, $\rkvar$ variables 
$\rkvar$ range over tuples over $\vals$.

\mypara{\IMPSEM States.}
In the \IMPSEM semantics,
each state $\istate$ is either the special \emph{error state} $\ERROR$ 
or a map from program variables to values such that
every base variable is mapped to a value in $\vals$, and 
every \REFREL variable of arity $n$ is mapped either to 
a tuple in $\vals^n$ or to the special \emph{undefined} value $\bot$.
Let $\istates$ denote the set of all a \IMPSEM-program states. 

\mypara{Initial State.} 
The initial state $\istate_0$ of an \ilang program in the \IMPSEM semantics is a map in which
every base variable is mapped to a fixed value from $\vals$,
and every \REFREL variable is mapped to $\bot$.

\mypara{Transition Relation.}
The transition relation is defined using a $\ipost$ operator, 
which is identical to $\rpost$ in the \RELSEM semantics except 
for the tuple-get and tuple-set instructions.
Figure~\ref{fig:impsemantics} shows the operator $\ipost$ for get and set operations.
Again, $\ipost$ is lifted to a set of states in the natural way.
Notice that the program halts if a get instruction is executed with
an \emph{undefined} \REFREL variable, or an $\ASSUME(p)$ is executed in a state
that does not satisfy $p$.

\mypara{Safety.} 
Let $\prgm$ be the program $\LOOP\{ \instr_1 \CHOOSE \ldots \CHOOSE \instr_n \}$.
The set of \emph{\IMPSEM-reachable states} of $\prgm$, denoted $\ireach(\prgm)$ is 
defined by induction as:
$$
\begin{array}{ll}
\ireach(\prgm, 0)   & \defeq  \set{\istate_0}\\
\ireach(\prgm, m+1) & \defeq  \bigcup \set{\ipost(\ireach(\prgm, m), \instr_j) \mid 1 \leq j \leq n}\\
\ireach(\prgm)      & \defeq  \bigcup \set{\ireach(\prgm, m) \mid 0 \leq m}
\end{array}
$$
A program $\prgm$ is \emph{\IMPSEM-safe} if $\ERROR \not \in \ireach(\prgm)$.

\section{From Type Constraints to \ilang Programs} \label{sec:equiv}

In this section we formalize the translation from type constraints
into \ilang programs and prove that the constraints are satisfiable
if and only if the translated program is safe.

\subsection{Translation}\label{sec:translation}

\begin{figure}[t]
\figbegin
\begin{displaymath}
\begin{array}{rll}
\multicolumn{3}{l}{\mbox{\textbf{Refinement Type Translation}}}  \\[4pt]
\translate{\reftyp{\valu}{\typ}{p}}_{get}	
& \defeq & \HAVOC{\valu}; \\
&        & \ASSUME(p) \\[4pt]

\translate{\reftyp{\valu}{\typ}{p}}_{set} 	
& \defeq & \ASSERT(p) \\[4pt]

\translate{\reftyp{\valu}{\typ}{\SUBST{\kvar}{x_1 \ldots x_n}{y_1 \ldots y_n}}}_{get} 
& \defeq & \GET{\rkvar}{(t_0,\ldots,t_n)}; \\
&	 & \ASSUME(y_1 = t_1); \\
&	 & \vdots \\
&	 & \ASSUME(y_n = t_n); \\
&	 & \ASSIGN{\valu}{t_0} \\[4pt]

\translate{\reftyp{\valu}{\typ}{\SUBST{\kvar}{x_1 \ldots x_n}{y_1 \ldots y_n}}}_{set} 
& \defeq & \SET{\rkvar}{(\valu,y_1,\ldots,y_n)} \\[8pt]

\multicolumn{3}{l}{\mbox{\textbf{Binding Translation}}}  \\[4pt]
\translate{\EXT{\ftyp{x}{\tliqs}}{\renv}} 
& \defeq & \translate{\tau}_{get};\ \ASSIGN{x}{\valu};\ \translate{\renv} \\[4pt]

\translate{\cdot} 
& \defeq & \SKIP \\[8pt]

\multicolumn{3}{l}{\mbox{\textbf{Constraint Translation}}}  \\[4pt]
\translate{\renv \deriv \tliqs_1 \subt \tliqs_2}
& \defeq & \translate{\renv};\ \translate{\tliqs_1}_{get};\ \translate{\tliqs_2}_{set} \\[8pt]

\multicolumn{3}{l}{\mbox{\textbf{Constraint Set Translation}}}  \\[4pt]
\translate{\set{c_1,\ldots,c_n}} 
& \defeq & \LOOP \{ \translate{c_1} \CHOICE \ldots \CHOICE \translate{c_n} \} \\
\end{array}
\end{displaymath}
\figend
\caption{\textbf{Translating Constraints to \ilang Programs}}
\label{fig:translate}
\end{figure}

Figure~\ref{fig:translate} formalizes the translation 
from (a set of) refinement type constraints $C$ to an \ilang program $\translate{C}$.
We use the WF constraints to translate each \REFREL 
variable $\kvar$ of arity $n+1$ into a corresponding 
tuple variable $\rkvar$ of arity $n+1$.

The translation is syntax-driven.
We translate each subtyping constraint
$\renv \deriv \tliqs_1 \subt \tliqs_2$ 
into a straight-line block of instructions with three parts: 
a sequence of instructions that establishes 
the environment bindings
($\translate{\renv}$),
a sequence of instructions that ``gets" the 
values corresponding to the LHS 
($\translate{\tliqs_1}_{get}$)
and a sequence of instructions that ``sets" the (LHS) values
into the appropriate RHS
($\translate{\tliqs_2}_{set}$).
The translation for a set of constraints is an infinite loop
that non-deterministically chooses among the blocks for each constraint.

Each environment binding gets translated as a ``get". 
Bindings with unknown refinements are translated into tuple-get operations,
followed by $\ASSUME$ statements that establish the equalities
corresponding to the pending substitutions.
Bindings with known refinements are translated into non-deterministic assignments
followed by a $\ASSUME$ that enforces that the refinement holds on the
non-deterministic value.

Each ``set" operation to an unknown refinement is translated into a
tuple-set instruction that writes the tuple corresponding to the pending
substitutions into the translated tuple variable.
Finally, each ``set" operation corresponding to a known refinement is
translated to an $\ASSERT$ instruction; intuitively, in such constraints 
the RHS defines an upper bound on the set of values populating the type, 
and the $\ASSERT$ serves to enforce the upper bound requirement in the
translated program.


The correctness of the procedure is stated by the following theorem.

\begin{theorem}{}\label{th:translate}
$C$ is satisfiable iff $\translate{C}$ is \emph{\RELSEM-safe}.
\end{theorem}

The proof of this theorem follows from the properties of the 
following function $\Convert$ that maps a set $\hat{\rstates}\subseteq\rstates$
of \RELSEM-states to constraint solutions:
$$\Convert(\hat{\rstates}) \defeq 
\lambda \kvar. \bigcup \set{\rstate(\rkvar) \mid \rstate \in \hat{\rstates}}$$
The function $\Convert$ enjoys the following property, which can be
proven by induction on the construction of $\rreach$, that relates the
satisfying solutions of the constraints to the \RELSEM-reachable states 
of the translated program. 
Theorem~\ref{th:translate} follows from the following observations.
%
If $\Soln$ satisfies $C$ then 
      $\Convert(\rreach(\translate{C}))(\kvar) \subseteq \Soln(\kvar)$
for all $\kvar$.
If $\ERROR \not \in \rreach(\translate{C})$ then
      $\Convert(\rreach(\translate{C}))$ satisfies $C$.

\subsection{Read-Write-Once Programs}
\label{sec:rwo}

At this point, via Theorem~\ref{th:translate}, we have reduced 
checking satisfiability of type constraints to the problem of
verifying assertions of \ilang programs under the 
(non-standard) \RELSEM semantics.
Unfortunately, under these semantics, the program contains variables 
($\rkvar$) which range over \emph{sets} of tuples. 
This makes it inconvenient to directly apply abstract-interpretation
based techniques for imperative programs which typically assume the
(standard) \IMPSEM semantics; each technique has to be painstakingly
adapted to the non-standard semantics.

We would be home and dry if we could prove the equivalence of the 
\RELSEM and \IMPSEM semantics; that is, if we could show that 
an \ilang program was \RELSEM-safe if and only if it was \IMPSEM safe.
Unfortunately, this is not true.

\myexsh 
Consider the \ilang program:
$$
\begin{array}{rlcll}
\LOOP\{ \quad 	& 
\begin{array}{l}
	\HAVOC{\valu};\\  
	\SET{\kvar}{(\valu)}
\end{array}
& 
\CHOICE		& 
\begin{array}{l}
                 \GET{\kvar}{(t_0)};\\
		 \ASSIGN{\valu}{t_0}; \ASSIGN{x}{\valu}; \\
		 \GET{\kvar}{(t_0)};\\
		 \ASSIGN{\valu}{t_0}; \ASSIGN{y}{\valu}; \\
		 \ASSERT{(x=y)}
\end{array}& \}
\end{array}
$$
This program \emph{is not} \RELSEM-safe as the set-operation in the first
instruction populates $\rkvar$ with the set of all integers,
and the get-operation in the second instruction can assign different values
to integer values to $x$ and $y$.
However the program \emph{is} \IMPSEM-safe as whenever the second
instruction executes, $\rkvar$ will be undefined or contain 
some arbitrary integer that is assigned to both $x$ and $y$, 
which causes the assert to succeed.

This example pinpoints exactly why the two semantics differ. 
In the \RELSEM semantics, in any given loop iteration, 
different gets on the same $\rkvar$ can return 
\emph{different} tuples, while in the \IMPSEM 
semantics the gets are correlated and return the same tuple.

\mypara{Read-Write-Once Programs.} 
An \ilang instruction is a \emph{read-write-once} instruction if
any \REFREL variable $\rkvar$ is read from and written 
to at most once in the instruction. 
That is, read-write-once means at most one write and at most one read
(and not at most one read or write).
An \ilang program is a \emph{read-write-once} program if each instruction 
in its loop is a read-write-once instruction.
We can show that for Read-Write-Once \ilang programs 
the \RELSEM and \IMPSEM semantics are equivalent.

\begin{theorem}{}\label{th:rwo-equiv}
If $\prgm$ is a \emph{read-write-once} $\ilang$ program then   
$\prgm$ is \emph{\RELSEM-safe} iff $\prgm$ is \emph{\IMPSEM-safe}.
\end{theorem}

To prove this theorem, we formalize the connection between the
reachable states under the two different semantics, using the function
$\EXPAND$, which maps a \RELSEM-state to a set of \IMPSEM states:
\begin{align*}
\EXPAND(\rstate) \defeq & 
	\left\{ \istate \mid 
		\begin{array}{ll}
                \istate(x) = \rstate(x) 	& \mbox{for base variables}\\
		\istate(\rkvar) = \mybar{v} 	& \mbox{if }\mybar{v} \in \rstate(\kvar)\\
		\istate(\rkvar) = \bot		& \mbox{if }\rstate(\rkvar) = \emptyset\\
		\istate = \ERROR		& \mbox{if }\rstate = \ERROR
                \end{array} 
        \right\} \\
\intertext{We lift the function to sets of \RELSEM states in the natural way:} 
\EXPAND(\hat{\rstates}) \defeq & \bigcup \set{\EXPAND(\rstate) \mid \rstate \in \hat{\rstates}}
\end{align*}
Next, we can show that read-write-once instructions enjoy the following
property, by case splitting on the form of $I$.

\begin{lemma}{\textbf{[Step]}}\label{lemma:step-lemma}
If $\instr$ is a read-write-once instruction then
$\EXPAND(\rpost(\rstate, \instr)) = \ipost(\EXPAND(\rstate), \instr)$.
\end{lemma}


We use this property to show that the reachable states under the different
semantics are equivalent.

\begin{lemma}{}\label{lemma:expand}
If $\prgm = \LOOP \set{\instr_1 \CHOICE \ldots \CHOICE \instr_n}$ 
is a read-write-once program, then $\EXPAND(\rreach(\prgm)) = \ireach(\prgm)$.
\end{lemma}
\includeProof{
\begin{proof}
To prove that $\ireach(\prgm) \subseteq \EXPAND(\rreach(\prgm))$, 
we show 
$${\forall m:\; \ireach(\prgm, m) \subseteq \EXPAND(\rreach(\prgm))}$$
by straightforward induction on $m$, noting that
$\istate_0 \in \EXPAND(\rstate_0)$, and 
$\ipost(\EXPAND(\rstate),\instr)\subseteq \rpost(\rstate,\instr)$ for
any \RELSEM-state $\rstate\in\rstates$, instruction $\instr$, and any
program $\prgm$ (not necessarily read-write-once).

To show inclusion in the other direction, 
we prove
$${\forall m:\; \EXPAND(\rreach(\prgm, m)) \subseteq \ireach(\prgm)}$$ 
by induction on $m$.
For the base case, 
$$\EXPAND(\rreach(\prgm,0)) = \ireach(\prgm, 0) \subseteq \ireach(\prgm)$$
by the definition of the initial states.
By induction, assume that 
$$\EXPAND(\rreach(\prgm, m)) \subseteq \ireach(\prgm)$$ 
Let ${\istate' \in \EXPAND(\rreach(\prgm, m+1))}$. 
By Lemma~\ref{lemma:step-lemma}, either 
$\istate'$ is already in $\rreach(\prgm, m)$, 
in which case the inductive hypothesis applies and 
hence $\istate' \in \ireach(\prgm)$, or
$$\istate' \in \ipost(\EXPAND(\rreach(\prgm, m), \instr_j)$$ 
for some $j$. That is, there is a 
$\istate \in \EXPAND(\rreach(\prgm, m)$ such that 
$\istate' \in \ipost(\istate, \instr_j)$.
From the induction hypothesis 
$\istate \in \ireach(\prgm)$. 
As $\ireach(\prgm)$ is closed under $\ipost$, 
we conclude $\istate' \in \ireach(\prgm)$.
\end{proof}
}


\subsection{Cloning}
\label{sec:cloning}

At this point, we have shown that the \IMPSEM semantics of
read-write-once programs are equivalent to the \RELSEM semantics.
All that remains is to show that the translation procedure of
Figure~\ref{fig:translate} produces read-write-once programs.
Unfortunately, this is not true.

\myexsh
Consider the following constraints:
$$
\begin{array}{c}
\emptyset \deriv \sreftyp{\kvar}\ ,
\emptyset  \deriv \sreftyp{\true} \subt \sreftyp{\kvar}\ , 
\EXT{\ftyp{x}{{\kvar}}}{\ftyp{y}{{\kvar}}}  \deriv \sreftyp{\true} \subt \sreftyp{x = y}
\end{array}
$$
It is easy to check that on the above constraints, the translation
procedure yields the \ilang program from the previous example, which is not
read-write-once.

The reason the translated program is not a read-write-once program is that 
there can be constraints $\renv \deriv \tliqs_1 \subt \tliqs_2$ 
in which $\kvar$ occurs in multiple places within $\renv$ and $\tliqs_1$.

To solve this problem, we can simply \emph{clone} the $\kvar$ variables 
that occur multiple times inside a constraint, and use different clones 
at each occurrence!
We formalize this as a procedure $\CLONE$ that maps a finite set of 
constraints to another finite set. The procedure works as follows. 
For each $\kvar$ that is read upto $n$ times in some constraint, we 
make $n$ clones, $\kvar^1,\ldots,\kvar^n$, and 
\begin{enumerate}
\item for the $i^{th}$ occurence of $\kvar$ within any constraint, 
    we use the $i^{th}$ clone $\kvar^i$ (instead of $\kvar$), and,
\item for each constraint where $\kvar$ appears on the right hand side, 
    we make $n$ clones of the constraints where in the $i^{th}$ cloned constraint, 
    we use $\kvar^i$ (instead of $\kvar$). 
\end{enumerate}
The first step ensures that each $\kvar$ is read-once in any constraint, and
the second step ensures that the clones correspond to exactly the same set of tuples 
as the original variable $\kvar$.
We can prove that $\CLONE$ enjoys the following properties.

\begin{theorem}{}\label{th:clone}
Let $C$ be a finite set of constraints.
\begin{enumerate}
\item $\translate{\CLONE(C)}$ is a read-write-once program.
\item $\CLONE(C)$ is satisfiable iff $C$ is satisfiable.
\end{enumerate}
\end{theorem}

It is easy to verify that $\translate{\CLONE(C)}$ is a read-write-once
program. 
Furthermore, any satisfying solution for the original
constraints can be mapped directly to a solution for 
the cloned constraints. 
To go in the other direction, we must map a solution that satisfies the
cloned constraints to one that satisfies the original constraints.
This is trivial if the solution for the cloned constraints 
maps each clone $\kvar^i$ to the same set of tuples.
We show that if the cloned constraints have a satisfying solution,
they have a solution that satisfies the above property.
To this end, we prove the following lemma that states 
that for \emph{any} set of constraints, the satisfying 
solutions are closed under intersection.

\begin{lemma}{}\label{lemma:solnintersect}
If $\Soln_1$ and $\Soln_2$ are solutions that satisfy $C$ 
then $\Soln_1 \cap \Soln_2 \defeq \lambda \kvar. \Soln_1(\kvar) \cap \Soln_2(\kvar)$
satisfies $C$.
\end{lemma}

Thus if $\Soln$ satisfies the cloned constraints
then by symmetry and Lemma~\ref{lemma:solnintersect} 
the solution that maps \emph{each} cloned variable 
to $\cap_{i=1}^{n}\Soln(\kvar^i)$ also satisfies 
the cloned constraints, and hence, directly yields 
a solution to the original constraints.

Finally, as a corollary of 
Theorems~\ref{th:translate},\ref{th:rwo-equiv},\ref{th:clone} 
we get our main result that reduces the question 
of refinement type constraint satisfaction,
to that of safety verification.

\begin{theorem}{}\label{th:equiv}
$C$ is satisfiable iff $\translate{\CLONE(C)}$ is \IMPSEM-safe.
\end{theorem}

While we state Theorems \ref{th:translate} and \ref{th:clone} as
preserving satisfiability, the proof shows how the solutions can be
effectively mapped between $C$ and $\translate{C}$ (or
$\translate{\CLONE(C)}$. 
In particular, while the intersection of two non-trivial solutions can
be a trivial solution, it would be guaranteed that in that case, the
trivial solution satisfies~$C$. 
Stated in terms of invariants, Lemma \ref{lemma:solnintersect} states
the observation that that there may be several non-comparable
inductive invariants to prove a safety property, but in that case, the
intersection of all the inductive invariants is also an inductive
invariant.


\section{Experiments}\label{sec:experiments}

\newcommand{\invpage}[1]{
  \begin{minipage}[h]{.35\linewidth}
    \begin{displaymath}
      \begin{array}{c}
        #1\\[\jot]
      \end{array}
    \end{displaymath}
  \end{minipage}
}
\newcommand{\typepage}[1]{
  \begin{minipage}[h]{.4\linewidth}
    \begin{displaymath}
      \begin{array}{c@{\;\defeq\;}l}
        #1\\[\jot]
      \end{array}
    \end{displaymath}
  \end{minipage}
}

\newcommand{\mathpage}[1]{
  \begin{minipage}[h]{.3\linewidth}
    \begin{displaymath}
      #1\\[\jot]
    \end{displaymath}
  \end{minipage}
}

\begin{table}[t]
\begin{small}
  \centering
  \begin{tabular}{|l||c|c|}
    \hline
    \textbf{Program} & \textbf{Time} & \textbf{Invariant} \\ \cline{3-3}
    & \textbf{(sec)}& \textbf{Refinement Types} \\ \hline\hline
    max & 0.091 & $\rkvar_1.1\leq\rkvar_1.0 \wedge
      \rkvar_1.2\leq\rkvar_1.0$ \\ \cline{3-3}
    & & $\rkvar_x \defeq \true, \rkvar_y \defeq \true,  \rkvar_1
    \defeq x \leq v \mathrel{\wedge} y \leq v$ \\ \hline
    sum & 0.071 & 
    $0 \leq \rkvar_2.0 \wedge \rkvar_2.1 \leq \rkvar_2.0 $ \\\cline{3-3}
    & & $\rkvar_k \defeq \true, \rkvar_2 \defeq 0 \leq v \wedge k \leq
    v$ \\ \hline
    foldn & 0.060 & 
    $0 \leq \rkvar_i.0 \wedge 0 \leq \rkvar_3.0 \wedge \rkvar_3.0 <
    \rkvar_3.2$ \\ \cline{3-3} 
    &&  $\rkvar_i \defeq 0 \leq v, \rkvar_3 \defeq 0 \leq v \wedge v <
    n$ \\ \hline
    arraymax & 0.135 & 
    $ 0\leq \rkvar_4.0 \wedge 0\leq \rkvar_5.0 \mathrel{\wedge} $\\
    & & $0\leq \rkvar_6.0 \wedge \rkvar_g.0 <
    \mathtt{len}(\rkvar_g.1)$ \\ \cline{3-3}
    & & $\rkvar_4 0 \defeq \leq v, \rkvar_5 \defeq 0 \leq v,$\\
    & & $\rkvar_6 \defeq 0 \leq v, \rkvar_g \defeq v < \mathtt{len(a)}$ \\\hline
    mask & 0.098 &
    $\rkvar_1.0 < \mathtt{len}(\rkvar_1.4) \wedge \rkvar_1.1 \leq \rkvar_1.0 \mathrel{\wedge}$ \\
    & & $0 \leq \rkvar_2.0 \wedge \rkvar_2.0 < \mathtt{len}(\rkvar_2.3)$\\\cline{3-3}
    & & $\rkvar_1 v < \mathtt{len(xs)} \wedge i \leq v,$\\
    & & $\rkvar_2 \defeq 0 \leq v \wedge v < \mathtt{len(a)}$ \\\hline
    samples & 0.117 &
    $ 0 \leq \rkvar_2.0 \wedge \rkvar_2.0 < \mathtt{len}(\rkvar_2.4) \mathrel{\wedge}$\\
    & & $0 \leq \rkvar_3.0 \wedge \rkvar_3.0 <
    \mathtt{len}(\rkvar_3.3) \mathrel{\wedge} 0 \leq \rkvar_6.0$ \\\cline{3-3}
    & & $\rkvar_2 \defeq 0 \leq v \wedge v < \mathtt{len(b)}, $ \\
    & & $\rkvar_3 \defeq 0 \leq v \wedge v < \mathtt{len(a)}, \rkvar_6 \defeq 0 \leq v$\\\hline
  \end{tabular}
  \caption{Experimental evaluation using a predicate abstraction-based
    verification tool on examples from~\cite{LiquidPLDI08}.
    The third column presents the invariant for the translated
program, and the resulting refinement types.}
  \label{tab:experiments}
\end{small}
\end{table}






We have implemented a verification tool for \ocaml programs based on \HMC.
We use the liquid types infrastructure implemented in \dsolve \cite{LiquidPLDI08}
to generate refinement type constraints from \ocaml programs.
We use \ARMC \cite{PADL07},
a software model checker using predicate abstraction and interpolation-based
refinement, as the verifier for the translated imperative program.

Table~\ref{tab:experiments} shows the results of running our tool on a
suite of small \ocaml examples from~\cite{LiquidPLDI08}.
For array manipulating programs, the safety objective is to prove
array accesses are within bounds.
For \textsc{max} we prove that the output is larger than input values.
For \textsc{sum} we prove that the sum is larger than the largest
summation term.

\begin{table}[t]
\begin{small}
  \centering
  \begin{tabular}{|l||c|c|c|}
   \hline
    \textbf{Program} & \textbf{Time} & \textbf{\# iterations}
    & \textbf{\# predicates}\\ \hline\hline
boolflip.ml & 2.17s & 7 & 21 \\ \hline
sum.ml & 0.24s & 5 & 14 \\ \hline
sum-acm.ml & 0.11s & 1 & 3 \\ \hline
sum-all.ml & 3.51s & 10 & 26 \\ \hline
mult.ml & 4.67s & 10 & 25 \\ \hline
mult-cps.ml & 780.24s & 11 & 27 \\ \hline
mult-all.ml & 18.44s & 9 & 24 \\ \hline\hline
boolflip-e.ml & 0.65s & \multicolumn{2}{c|}{} \\ \cline{1-2}
 sum-e.ml & 0.01s & \multicolumn{2}{c|}{} \\ \cline{1-2}
sum-acm-e.ml & 0.02s & \multicolumn{2}{c|}{} \\ \cline{1-2}
sum-all-e.ml & 0.79s & \multicolumn{2}{c|}{} \\ \cline{1-2}
mult-e.ml & 0.01s & \multicolumn{2}{c|}{} \\ \cline{1-2}
mult-cps-e.ml & 7.69s & \multicolumn{2}{c|}{} \\ \cline{1-2}
mult-all-e.ml & 144.93s & \multicolumn{2}{c|}{} \\ \hline
\end{tabular}
  \caption{Experimental evaluation of our tool on Depcegar
    benchmarks~\cite{TerauchiPOPL2010}.
    The third column presents the number of abstraction refinment
    iterations required by \ARMC. 
    The last column gives the number of predicates discovered by
    \ARMC.
    For the programs with suffix ``-e'', which are incorrect, we omit
    the number of iterations and predicates and only show the time
    required by \ARMC to find a counterexample. }
  \label{tab-terauchi-popl2010}
\end{small}
\end{table}


Table~\ref{tab-terauchi-popl2010} presents the running time of our
tool on the benchmark programs for the Depcegar
verifier~\cite{TerauchiPOPL2010}.
We observe that despite of our blackbox treatment of \ARMC as a
constraint solver we obtain competitive running times compared to
Depcegar on most of the examples (Depcegar uses a customized procedure
for unfolding constraints and creating interpolation queries that
yield refinement types).

Most of the predicates discovered by the interpolation-based
abstraction refinement procedure implemented in \ARMC fall into the
fragment ``two variables per inequality.''
The example \textsc{mask} required a predicate that refers to three
variables, see~$\rkvar_1$.
While our initial experiments used a CEGAR-based tool, we expect optimized
abstract interpreters for numerical domains to also work well for this class of properties.


\section{Extensions and Related Work}\label{sec:discussion}

\subsection{Completeness}

The soundness of safety verification for higher-order programs 
for any domain follows from the soundness of constraint generation 
(\eg Theorem~1 in \cite{LiquidPLDI08}) and Theorem~\ref{th:equiv}. 
Since the safety verification problem for higher-order programs
is undecidable, the technique cannot be complete in general.
Even in the finite-state case, in which
each base type has a finite domain (\eg booleans), 
completeness depends on the generation of type constraints.
For example, in our examples and in our implementation, we have
assumed a \emph{context insensitive} constraint generation from program
syntax, \ie we have not distinguished the types of the same function
at different call points.
This entails a loss of information, as the following example demonstrates.
Consider
\begin{verbatim}
let check f x y = assert (f x = y) in
check (fun a -> a) false false ;
check (fun a -> not a) false true
\end{verbatim}
%
where the builtin function $\ASSERT$ has the type 
$\reftyp{\valu}{\ttbool}{\valu} \rightarrow \ttunit$.
The refinement template for \verb+check+ 
generated by our constraint generation process is 
\[
(\ttx: \reftyp{\valu}{\ttbool}{\kappa_1}\rightarrow \set{\kappa_2}) \rightarrow \set{\kappa_3} \rightarrow \set{\kappa_4}\rightarrow \ttunit
\]
which is too weak to show that the program is safe.
This is because the template ``merges'' the two call sites for \verb+check+.

One way to get context sensitivity is through \emph{intersection types} 
\cite{FreemanPfenning91,Dunfield,NaikPalsberg,KobayashiPOPL09}.
For the above example, we can show type safety 
using the following refined type for \verb+check+:
\[
\begin{array}{cl}
\bigwedge & \begin{array}{l}
            (\ttx: \ttbool \rightarrow \set{\valu = \ttx})\rightarrow \set{\lnot \valu}\rightarrow \set{\lnot\valu}\rightarrow \ttunit\\
            (\ttx: \ttbool \rightarrow \set{\valu = \lnot\ttx})\rightarrow \set{\lnot \valu}\rightarrow \set{\valu}\rightarrow \ttunit
            \end{array}
\end{array}
\]
It is important to note that Theorems~\ref{th:translate} and~\ref{th:rwo-equiv} 
hold for \emph{any} set of constraints.
Thus, one way to get completeness in the finite state case 
is to generate refinement templates using intersection types, 
perform the translation to \ilang programs, 
and then using a complete invariant generation 
technique for finite state systems.
The key observation (made in \cite{KobayashiPOPL09}) 
that ensures a finite number of constraints, is 
that there is at most a finite number of ``contexts'' in the finte state case,
and hence a finite number of terms in the intersection types.
The bad news is that the bound on the number of contexts is 
$\mathsf{exp}_n(k)$, where $n$ is the highest order of any 
function in the program, $k$ is the maximum arity of any function in the program,
and $\mathsf{exp}_n(k)$ is a stack of $n$ exponentials, defined by 
$\mathsf{exp}_0(k) = k$, and $\mathsf{exp}_{n+1}(k) = 2^{\mathsf{exp}_n(k)}$.

Fully context-sensitive constraints are used in \cite{KobayashiPOPL09}
to show completeness in the finite case, at the price of
$\mathsf{exp}_n(k)$ in {\em every case}, not just the worst case.
In our exposition and our implementation, we have traded 
off precision for scalability: while we lose precision 
by generating context-insensitive constraints, we avoid 
the $\mathsf{exp}_n$ blow-up that comes with full context sensitivity.
However, it has been shown through practical benchmarks that since the types themselves capture
relations between the inputs and outputs, the context-insensitive
constraint generation suffices to prove a variety of complex programs safe
\cite{LiquidPLDI08, LiquidPLDI09, GordonRefinement09}.

When considering completeness properties in special cases, we point
out completeness wrt.~the discovery of refinement predicates in
octagons/difference bounds abstract domains~\cite{MineOctagon06} and
template-based invariant generation for linear
arithmetic~\cite{ColonCAV03} and extensions with uninterpreted
function symbols~\cite{BeyerVMCAI07}, which carries over from
respective verification approaches.

\subsection{Related Work}

\mypara{Higher-Order Programs.}
Kobayashi \cite{KobayashiPOPL09,KobayashiLICS09} gives an algorithm
for model checking arbitrary $\mu$-calculus properties of finite-data
programs with higher order functions by a reduction to model checking
for higher-order recursion schemes (HORS)~\cite{Ong}.
For safety verification, \HMC shows a promising alternative.

First, the reduction to HORS critically depends on a finite-state abstraction of the data.
In contrast, our reduction defers the data abstraction to the abstract interpreter working
on the imperative program, thus enabling the direct application of abstract interpreters working
over infinite domains. 
Since abstract interpreters over infinite abstract domains are strictly 
more powerful than (infinite families of) finite ones \cite{CousotCousot92comparison}, 
our approach can be strictly more powerful for infinite-state programs.

Second, in the translation of an abstracted program to a HORS,
this algorithm eliminates Boolean variables by enumerating 
all possible assignments to them, giving an exponential
blow-up from the program to the HORS. 
In contrast, our technique preserves the Boolean state \emph{symbolically}, 
enabling the use of efficient symbolic algorithms for verification.
For example, for the simple example:
\begin{verbatim}
let f b1 ... bn x = 
  if (b1 || ... || bn) then lock x;
  if (b1 || ... || bn) then unlock x 
in let f (*) ... (*) (newlock ()) 
\end{verbatim}
where we wish to prove that lock and unlock alternate. 
Kobayashi's translation \cite{KobayashiPOPL09} gives an {\em exponential} sized HORS,
with a version of $\ttf$ for each assignment to \verb+b1,...,bn+.
In contrast, our reduction preserves the source-level expressions and is linear, 
and amenable to symbolic verification techniques (e.g., BDDs).
Previous experience with software model checking \cite{SLAMPOPL02,HJMM04,fsoft06} 
shows that the number of reachable states is often drastically 
smaller than $2^p$ where $p$ is the number of Booleans.  
Thus, the pre-processing step that enumerates Booleans 
may not lead to a scalable implementation.

Might \cite{Might07} describes {\em logic-flow analysis}, a general safety verification 
algorithm for higher-order languages, which is the product 
of a $k$-CFA like call-strings analysis and a form of SMT-based
predicate abstraction (together with widening).
In contrast, our work shows how higher-order
languages can be analyzed directly via 
abstract analyses designed for first-order imperative languages.

Inference of refinement types using conterexample-guided techniques
was recentrly identified as a promising
direction~\cite{UnnoPPDP09,TerauchiPOPL2010}.
In contrast, our approach is not limited to CEGAR and facilitates the
applicability of a wide range abstract interpretation techniques for
precise reasoning about program data.

\mypara{Software Verification.}
This work was motivated by the recent success in 
software model checking for first-order imperative
programs \cite{SLAMPOPL02,HJMM04,CousotPLDI03,McMillan06}, 
and the desire to apply similar techniques to modern programming
languages with higher order functions.
Our starting point was refinement types \cite{FreemanPfenning91,Knowles07},
implemented in dependent ML \cite{XiPfenning99} to give strong static guarantees, 
and the work on liquid types \cite{LiquidPLDI08,LiquidPLDI09}
that applied predicate abstraction to infer refinement types.
By enabling the application of automatic invariant generation from software
model checking,
\HMC reduces the need for programmer annotations in refinement type systems.

\bibliographystyle{plain}
\bibliography{sw}

\end{document}